\documentclass[remotesensing,article,accept,moreauthors,pdftex]{Definitions/mdpi}
\usepackage{gensymb}
\usepackage{amsmath}
\usepackage{amssymb}
\usepackage{epsfig}
\usepackage{graphicx}
\usepackage{subfigure}
\usepackage{epstopdf}
\usepackage{setspace}
\usepackage{booktabs}
\usepackage{multirow}
\usepackage{makecell}
\usepackage{color}
\usepackage{xpatch}
\usepackage{subfigure}
\makeatletter
\renewcommand{\@thesubfigure}{\normalsize(\textbf{\alph{subfigure}})}
\makeatother

\makeatletter
\def\changeBibColor#1{%
 \in@{#1}{tait2005introduction,clemente2017automatic,eryildirim2009man,sun2016sar}% list of colored bib items
 \ifin@\color{black}\else\normalcolor\fi
}

\xpatchcmd\@bibitem
 {\item}
 {\changeBibColor{#1}\item}
 {}{\fail}

\xpatchcmd\@lbibitem
 {\item}
 {\changeBibColor{#2}\item}
 {}{\fail}
\makeatother
\firstpage{1}
\pubvolume{12}
\issuenum{23}
\articlenumber{3863}
\pubyear{2020}
\copyrightyear{2020}
\history{Received: 13 October 2020; Accepted: 22 November 2020; Published: date}
\Title{When Deep Learning Meets Multi-Task Learning in SAR ATR: Simultaneous Target Recognition and Segmentation}

\Author{Chenwei Wang, Jifang Pei *, Zhiyong Wang, Yulin Huang, Junjie Wu, Haiguang Yang and \\Jianyu Yang} %Please carefully check the accuracy of names and affiliations.

\AuthorNames{Chenwei Wang, Jifang Pei, Zhiyong Wang, Yulin Huang, Junjie Wu, Haiguang Yang and Jianyu Yang}

\address[1]{%
The Department of Electrical Engineering, University of Electronic Science and Technology of China, Chengdu 611731, China; 201811011927@std.uestc.edu.cn (C.W.); 201922011020@std.uestc.edu.cn (Z.W.); yulinhuang@uestc.edu.cn (Y.H.); junjie\_wu@uestc.edu.cn (J.W.); yanghaiguang@uestc.edu.cn (H.Y.); jyyang@uestc.edu.cn (J.Y.)}

\corres{\hangafter=1 \hangindent=1.05em \hspace{-0.82em}Correspondence: jifangpei@uestc.edu.cn}
\abstract{With the recent advances of deep learning, automatic target recognition (ATR) of synthetic aperture radar (SAR) has achieved superior performance. By not being  limited to the target category, the SAR ATR system could benefit from the simultaneous extraction of multifarious target attributes. In this paper, we propose a new multi-task learning approach for SAR ATR, which could obtain the accurate category and precise shape of the targets simultaneously. By introducing deep learning theory into multi-task learning, we first propose a novel multi-task deep learning framework with two main structures: encoder and decoder. The  encoder is constructed to extract sufficient image features in different scales for the decoder, while  the decoder is a tasks-specific structure which   employs these extracted features adaptively and optimally to meet the different feature demands of the recognition and segmentation. Therefore, the proposed framework has the  ability to achieve   superior recognition and segmentation performance. Based on the Moving and Stationary Target Acquisition and Recognition (MSTAR) dataset, experimental results   show  the superiority of the proposed framework in   terms  of recognition and segmentation.}

 % List three to ten pertinent keywords specific to the article, yet reasonably common within the subject discipline.

\begin{document}
%%%%%%%%%%%%%%%%%%%%%%%%%%%%%%%%%%%%%%%%%%

%%%%%%%%%%%%%%%%%%%%%%%%%%%%%%%%%%%%%%%%%%

\section{Introduction}
Synthetic aperture radar (SAR) is an important microwave remote sensing system in the domains of military and civilian applications \cite{wang2019parking,wang2021deep}. With the high-resolution coherent imaging capability of all weather and all day  penetration, it can obtain more distinct information than optical sensors, infrared sensors,   etc. \cite{curlander1991synthetic,moreira2013tutorial}. Moreover, it is able to acquire abundant backscattering characteristics of the targets. These backscattering characteristics contain unique identifying information of target attributes, which is often difficult to accurately interpret from the perspective of human vision. Besides, it is usually a hard task to accomplish real-time processing when the size and number of SAR images are increasing. Therefore, SAR automatic target recognition (ATR) has become one of the most crucial and challenging issues in SAR application.

Basically, the fundamental problem of SAR ATR is to locate and recognize the objects of interest in an environment with clutters in SAR images \cite{dudgeon1993overview,wang2022sar,wang2021deep}. The standard architecture of the SAR ATR system proposed by MIT Lincoln Laboratory has three main stages: detection, discrimination   and   classification \cite{bhanu1986automatic}. In the detection stage, a constant false alarm rate (CFAR) detector is employed to localize where a potential target is likely to exist in the SAR image. Then, in the discrimination stage, some specific discriminating criteria are adopted to reject cultural and natural clutter false alarms. In the classification stage, an elaborate and efficient classifier provides additional false alarm rejection and categorizes the remaining detections as specific target types.

Many novel classification algorithms and systems have been proposed in recent years and performed well in applications \cite{zhao2001support,he2014forward,zhang2011joint}. These various methods for the classification stage in general can be taxonomized into two mainstream paradigms: template-based and model-based. The template-based taxon is a pattern recognition approach solely relying on templates to represent the targets \cite{novak1997automatic}. These templates could be two-dimensional target templates or extracted feature vectors. The process of the template-based taxon involves two distinctive stages: offline classifier training and online classification. Despite templated-based taxon's simplicity and popularity \cite{el2016automatic}, it may be unable  to cope with   extended operating conditions (EOCs). Unlike the template-based taxon, the model-based taxon mainly focuses on representing the characterization of the physical structure of the target \cite{ikeuchi1996model}. Typically, a model-based taxon consists of two main stages: holistic physical model construction for the target and the online classification prediction that yields close resemblance to the input SAR chips. Despite model-based taxon can circumvent the EOCs to some extent, it also faces the problem of the additional complexity in the SAR ATR system.

In recent decades, deep learning has been applied in signal and image processing fields and demonstrated its superior performance. As for the SAR ATR application,   many excellent studies have proposed many deep learning methods with outstanding results \cite{tait2005introduction,clemente2017automatic,eryildirim2009man,sun2016sar,bau2017network,wang2022recognition,wang2023entropy,wang2023sar,wang2022global,wang2022semi,wang2021multiview,wang2020multi}.   Chen et al. \cite{chen2016target} proposed an all-convolutional network replacing all the dense layers with the convolutional layers, which leads to outstanding recognition performance. Wagner et al. \cite{wagner2016sar} proposed a combination of a convolutional neural network and support vector machines to incorporate prior knowledge and enhance its robustness against imaging errors. Jiao et al. \cite{jiao2018densely} proposed a multi-scale and multi-scene ship detection approach for SAR images, which could detect small scale ships and avoid the interference of inshore complex background.   Li et al. \cite{li2020block} proposed block sparse Bayesian learning (BSBL) to synthetic aperture radar (SAR) target recognition, which considers the azimuthal sensitivity of SAR images and the sparse coefficients on the local dictionary.

However, most of the existing SAR ATR methods only focus on improving the detection or recognition performance and still need various separate subsystems for different functions. In practice, the whole system of SAR ATR has great demand in acquiring multifarious information of the given target, such as location, category, shape, morphological contour, ambient relationships,   etc. Furthermore, when multiple subsystems are employed to achieve the goals of detection, recognition,  etc., the complexity of the whole SAR ATR system will be too high to meet the practical demand.

In practice, for analyzing the purpose of the detected targets, it is necessary to get enough target attributes which are composed of multi-dimensional information. For example, it is critical to gain the categories and the tracks of detected ships simultaneously to judge if they are going to transport or attack. The category and geometric structure of the target contain substantial information among the multifarious target attributes \cite{pei2014active}. Therefore, it is necessary and valuable to extract the category and geometric structure of the given target with one system, namely one SAR ATR system deals with multiple tasks simultaneously.

Fortunately, multi-task learning (MTL)  can handle  different related tasks simultaneously, which can refer to the joint learning of multiple problems, enforce a common intermediate parameterization and replace multiple subsystems with one system. With the relevance of the different tasks, it could improve the generalization performance of the system, which is caused by leveraging the domain-specific information contained in the training dataset of related tasks \cite{parameswaran2010large,ruder2017overview,evgeniou2004regularized}. Besides, these related tasks are learned simultaneously by extracting and utilizing appropriate shared information across tasks. From a machine learning point of view, MTL could be regarded as a form of inductive transfer, which could improve a model by introducing an inductive bias provided by the related tasks \cite{zhang2012convex,lounici2009taking}.

Considering that the superior performance of deep learning, by introducing the deep learning theory into the framework of MTL, MTL will acquire the capability of adaptive feature learning and powerful feature representation to promote its performance \cite{lecun2015deep}, which would be a perfect encounter in SAR ATR. Furthermore, it is possible that a neural network MTL module can increase the performance of the whole SAR ATR system using the relevance between tasks.

Therefore, in this paper, we propose a novel multi-task deep learning framework for recognition and segmentation of the SAR target to obtain both its category and shape information. First, we construct a multi-task deep learning framework to complete target recognition and segmentation in SAR images, which consists of two parts: encoder and decoder. Second, a shared encoder is designed to extract effective features in different scales for morphological segmentation and recognition. Then, through constructing two different sub-network structures, the two decoders have the capability of employing these extracted features adaptively and optimally to meet the different feature demands of the recognition and segmentation. Therefore, the proposed multi-task  framework has the capability of extracting sufficient category and shape information of the SAR target.

The remainder of this paper is organized as follows. An overview of the multi-task deep learning framework is presented in Section \ref{5786313}. The specific design and instantiation of the proposed framework are given in Section \ref{764367}. Section 4 evaluates the performance of our proposed framework with experiments.   Section 5 gives the conclusions.
%%%%%%%%%%%%%%%%%%%%%%%%%%%%%%%%%%%%%%%%%%
\section{MTL Deep Learning Framework for SAR Target Segmentation and Recognition}
\label{5786313}

As mentioned above, the category and geometric structure of the targets are able to provide sufficient information of the SAR targets in practice. Therefore, we propose an MTL deep learning framework to efficiently extract multifarious attributes of the SAR target and achieve the recognition and segmentation simultaneously.

\begin{figure}[H]
\centering
\includegraphics[width=6.1in]{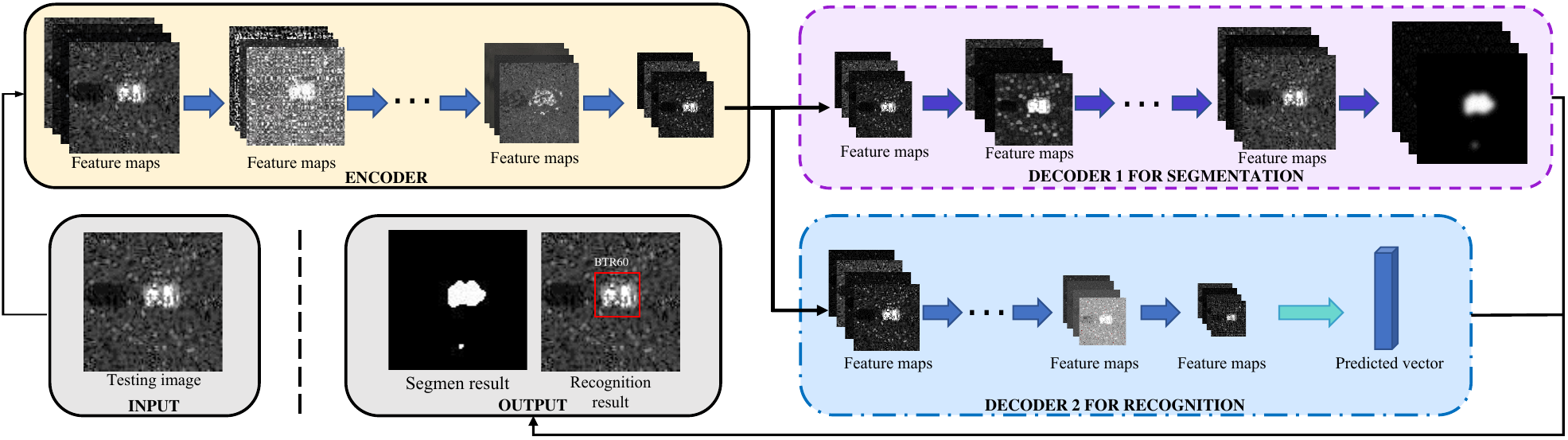}
\caption{Proposed MTL deep learning framework.}\label{framework}
\end{figure}

The proposed MTL deep learning framework mainly consists of two parts , as shown in Figure \ref{framework}: encoder and decoder. The encoder is a special structure which is utilized to extract optimal image feature from SAR image to achieve accuracy recognition and segmentation. The key point of the encoder construction is to provide sufficient image features in different scales for the decoder. Then, the decoder is a task-specific structure which is divided into two sub-decoders. The decoder for the precise segmentation is constructed to adopt the fusion of the extracted features in different scales. These features represent the overall contour and local details of the target. Meanwhile, the structure for recognition should finish further extraction and fusion of optimal image features to realize the accurate recognition of the targets. Through the above structures, the proposed multi-task  deep learning framework can extract optimal features layer by layer from SAR images and employ these extracted features adaptively and optimally to meet the different feature demands of the recognition and segmentation.

\section{Network Architecture of MTL Deep Learning Framework}
\label{764367}
In this section, a specific implementation of the proposed MTL deep learning framework and the details   of its configuration are presented. First, we elucidate the structure of the specific implementation. Then, the configurations of each layer    are  presented. Finally, the joint loss of the proposed network and the training implementation    are  given.

\subsection{Specific Implementation}
\begin{figure}[H]
\centering
\includegraphics[width=6.1in]{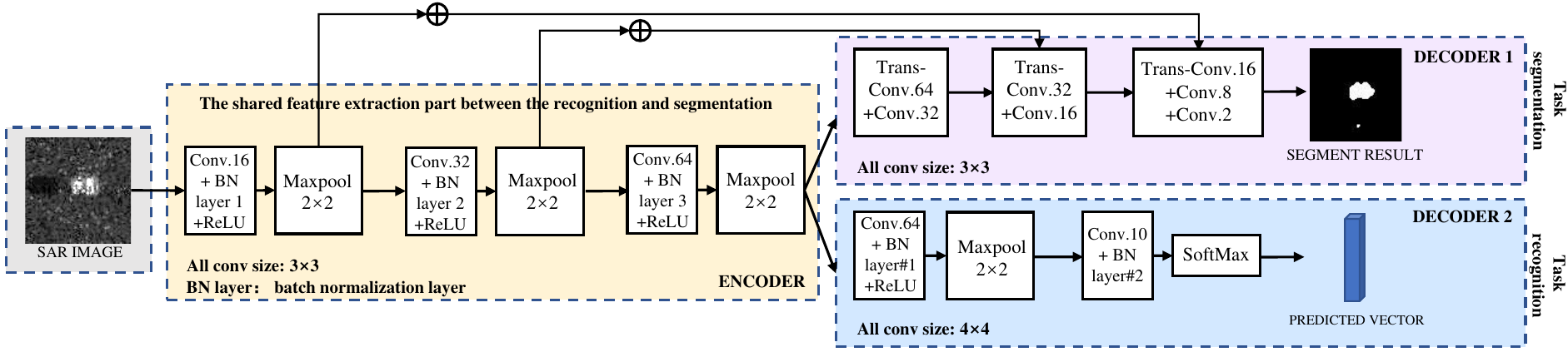}
\caption{Specific implementation of the proposed framework.}\label{network}
\end{figure}

The specific implementation of the proposed MTL deep learning framework is presented in Figure \ref{network}.

   To gain sufficient image information to achieve the recognition and segmentation of the targets, the encoder is designed to   consist of three convolutional layers and three max pooling layers to extract different forms of image features in different scales. A rectified linear unit (ReLU) \cite{nair2010rectified} is adopted as an activation function after each convolutional layer, which could increase nonlinear capability. A batch normalization \cite{ioffe2015batch} is adopted before each convolutional layer, which could make the middle data distribution more consistent with the distribution of the input data and ensure the nonlinear expression ability of the whole architecture. Therefore, the encoder gets the capability of fitting nonlinear data distribution and acquiring a different form of optimal image features.

Then, owing to the different demands of structure and image feature for the recognition and segmentation, the decoder is designed, respectively, for the two tasks, whose specific forms are two different sub-decoders with two different feature utilizations. The sub-decoder for the recognition consists of two convolutional layers and one max pooling layer. At the last convolutional layer, SoftMax is adopted as a classifier to get the normalized probability distribution of the recognition results. As for the segmentation, the sub-decoder is designed as the structure consist of three transposed convolutional layers \cite{zeiler2010deconvolutional} and three convolutional layers. After each convolutional layer, there is one skip connection \cite{zhang2018image} for being combined with the image features extracted by the encoder. There are no activation functions after each convolutional layer. Through the two specific structures of decoder for the recognition and segmentation of the targets, the decoder gets the capability of gaining accurate recognition and precise segmentation.

The details of those layers, activation functions, etc. are  described in the following.

\subsection{Convolutional Layer and Transposed Convolutional Layer}
The convolutional layer is the main component of the whole architecture to percept the local image information and extracts the image feature. Sparse connectivity and weight sharing are two advantages in the convolutional layer to reduce the number of parameters. Sparse connectivity means that the size of connection fields between the feature maps of the $(l-1){\mathop{\rm th}\nolimits}$ layers and the $l{\mathop{\rm th}\nolimits}$ convolutional layer is the same as the size of convolutional kernels. Weight sharing means that each convolutional kernel is employed to be calculated with all the spatial area in the convolutional layer. Given the $i{\mathop{\rm th}\nolimits}$ feature map in the $(l-1){\mathop{\rm th}\nolimits}$ layers as ${\bf{x}}_i^{l - 1}$, ${\bf{w}}_i^{l - 1}$ as one convolutional kernel and $b_i^{l - 1}$ as the bias in the $l{\mathop{\rm th}\nolimits}$ convolutional layer. The operation of the $l{\mathop{\rm th}\nolimits}$ convolutional layer can be presented as\vspace{6pt}
\begin{align}
\label{rangereso1}
{\bf{x}}_j^l = \sum\limits_{i = 1} {({\bf{x}}_i^{l - 1}) * ({\bf{w}}_{ij}^l)} + b_{j}^l
\end{align}
where $*$ denotes the convolution.

At the same time, the operation of the convolutional layers can be presented  as
\begin{align}
\label{rangereso1}
{{\bf{x}}^l} = {{\bf{W}}^l}{{\bf{x}}^{l - 1}} + {{\bf{b}}^l}
\end{align}
where ${{\bf{x}}^l}$ denotes the ${n^l}$ dimension output vector which   is reshaped into the matrix, ${{\bf{x}}^{l - 1}}$ denotes the ${n^{l - 1}}$ dimension input vector which is reshaped from the matrix and  ${{\bf{W}}^l}$ denoted the reshaped convolutional kernel whose size is ${n^l} \times {n^{l - 1}}$.

The transposed convolutional layer is an up-sampling method, which could seek for the optimal parameter to up-sample the images. The transposed convolutional layer is actually the reverse operation of the convolutional layer, which means that the forward and backward of the transposed convolutional layer are reverse to the convolutional layers. The operation of the transposed convolutional layer can be described as follows. First, the input image is padded with zero to expand the size. Then, the padded input images are convolved with the transposed convolutional kernels \cite{zeiler2010deconvolutional}. After each operation of convolution, the position of the next convolution is shifted  by the set stride.

The transposed convolutional layer is a main component of the decoder for the segmentation, which is adopted to up-sample and integrates the extracted feature maps adaptively layer by layer. The output size of the $l{\mathop{\rm th}\nolimits}$ transposed convolutional layer with the factor $s_T^l$ is equal to the convolutional layer with a fractional stride $\frac{1}{{s_T^l}}$. Given the $i{\mathop{\rm th}\nolimits}$ feature map in the $(l-1){\mathop{\rm th}\nolimits}$ layers as ${\bf{x}}_i^{l - 1}$, the operation of the transposed convolutional layer can  be  presented  as
\begin{align}
\label{rangereso1}
{{\bf{x}}^l}{\rm{ = }}{({{\bf{W}}^l})^T}{{\bf{x}}^{l - 1}} + {b^l}
\end{align}
where ${{\bf{W}}^l}$ denotes the reshaped convolutional kernel, whose size is ${n^l} \times {n^{l - 1}}$.

\subsection{Batch Normalization and Rectified Linear Unit}
Batch normalization is a trick to train a deep learning network. It not only can accelerate the convergence speed of the network, but also solve the problem called gradient dispersion to a certain extent, which makes it easier and more stable to train a deep learning network \cite{santurkar2018does}. The processing of batch normalization could be divided into three steps as following. First, given a batch of the input images as $B = \left\{ {{x_1},{x_2}, \ldots ,{x_m}} \right\}$, the average value and the variance of each training data batch are calculated by
\begin{align}
\label{rangereso1}
{\mu _B} = \frac{1}{m}\sum\limits_{i = 1}^m {{x_i}}
\end{align}
\begin{align}
\label{rangereso1}
\sigma _B^2 = \frac{1}{m}\sum\limits_{i = 1}^m {{{\left( {{x_i} - {\mu _B}} \right)}^2}}
\end{align}
where ${\mu _B}$ is the average value of this batch $B$ and  $\sigma _B^2$ is the variance. Then, the batch $B = \left\{ {{x_1},{x_2}, \ldots ,{x_m}} \right\}$ is normalized by ${\mu _B}$ and $\sigma _B^2$ to get the 0--1 distribution:
\begin{align}
\label{rangereso1}
{\hat x_i} = \frac{{{x_i} - {\mu _B}}}{{\sqrt {\sigma _B^2 + \varepsilon } }}
\end{align}
where $\varepsilon $ is a small positive number to avoid the divisor as zero. Finally, the normalized batch $B$ is subjected to  scale transformation and translation by
\begin{align}
\label{rangereso1}
B{N_{\gamma ,\beta }}\left( {{x_i}} \right) = \gamma {\hat x_i} + \beta
\end{align}
where $\gamma $ is the scale factor and $\beta $ is the translation factor. $B{N_{\gamma ,\beta }}\left( \cdot \right)$ is denoted as the operation of the batch normalization. The two learnable parameters, $\gamma $ and $\beta $, are introduced to solve the problem that the expression ability of the network is decreased, which is caused by the normalized batch being basically limited to the normal distribution \cite{cooijmans2016recurrent}.

The Rectified linear unit (ReLU) is an activation function which has less computational complexity than other activation functions \cite{xu2015empirical}, such as sigmoid,   and   solves  the problem called vanishing gradient to a certain extent. The formula of the ReLU can  be  presented  as
\begin{align}
\label{rangereso1}
f\left( {x_i^j} \right) = {\mathop{\rm ReLU}\nolimits} (x_i^j) = \left\{ {\begin{array}{*{20}{c}}
{x_i^j{\kern 1pt} {\kern 1pt} {\kern 1pt} {\kern 1pt} {\kern 1pt} {\kern 1pt} {\kern 1pt} if{\kern 1pt} x_i^j > 0}\\
{0{\kern 1pt} {\kern 1pt} {\kern 1pt} {\kern 1pt} {\kern 1pt} {\kern 1pt} {\kern 1pt} {\kern 1pt} if{\kern 1pt} x_i^j \le 0}
\end{array}} \right.
\end{align}
The ReLU will make the output of some feature maps   zero, which leads to the sparsity of the network and alleviates  the occurrence of the overfitting problem.
\subsection{Max Pooling and SoftMax}
The max pooling layer is utilized to integrate the information of the feature maps with reducing the number of parameters and the computational complexity of the whole network. The operation of the max pooling layer is to get the maximum value in the window of the feature maps as
\begin{align}
\label{rangereso1}
{p_i} = \mathop {\max }\limits_{\left( {u,v} \right) \in P} {x_i}\left( {u,v} \right)
\end{align}
where ${u,v}$ is the coordinate of the pixels in the pooling window, ${p_i}$ is the output of the max pooling layer and  $P$ is the pooling window. Although the max pooling layer has many advantages, it could also pool some crucial information for the segmentation or other tasks.

SoftMax is adopted as a classifier that could normalize the output of the network to be understood as posterior probability with the original intention to make the effect of the feature on probability multiplicative. Given the output vector of the network before SoftMax as ${{\bf{x}}^L}{\rm{ = }}\left\{ {x_1^L,x_2^L, \ldots ,x_C^L} \right\}$, the formula of SoftMax can  be  presented  as
\begin{align}
\label{rangereso1}
p\left( {{y_i}\left| {x^L} \right.} \right) = \frac{{\exp (x_i^L)}}{{\sum\limits_{k = 1}^C {\exp (x_k^L)} }}
\end{align}
where $C$ is the number for the target types, ${y_i}$ is the one-hot vector of the target type and $\exp \left( \cdot \right)$ is the power of $e$. Through the operation of SoftMax, the probability of each type of target is acquired corresponding to each element in the output vector of SoftMax.
\subsection{Joint Loss and Backpropagation}
The Joint loss is the combination of each task's loss, which could highly influence the performance of the whole framework. Through choosing the appropriate weights between each task's loss, the joint loss not only consider the difference between tasks, but also take the advantage of the relevance between tasks, which could lead to a better performance of the whole framework \cite{sun2014deep}. As for the target recognition and segmentation, the target recognition needs to utilize the features of the scattering distribution and target morphology, which is the same as the target segmentation \cite{badrinarayanan2017segnet}. Therefore, there is a strong coherence and relevance between the recognition and segmentation of target in the SAR image.

In the proposed multi-task  deep learning framework, the joint loss is set as the weighted sum of the recognition loss and the segmentation loss. The recognition loss is set as the cross-entropy cost function, which is presented as
\begin{align}
\label{rangereso1}
{L_r}\left( {{\bf{w,b}}} \right) = - \sum\limits_{i = 1}^C {{y_i}\log \left( {p\left( {{y_i}\left| {{{\bf{x}}^L}} \right.} \right)} \right)}
\end{align}

In nature, the target segmentation is a kind of classification in pixel level. To achieve accurate segmentation, the distance between the segmentation result and the ground truth should be calculated. Therefore, the segmentation loss is set as the cross-entropy cost function of all the pixels in a SAR chip.  The segmentation loss is averaged to the same unified scale as the recognition loss, which   leads to better and more robust performance \cite{nielsen2015neural}. The function of the segmentation loss is defined as
\begin{align}
\label{rangereso1}
{L_s}\left( {{\bf{w,b}}} \right) = - \frac{1}{{n}}\sum\limits_{i = 1}^V {{{\bf{s_i}}}\log \left( {{\bf{p}}\left( {{{\bf{s_i}}}\left| {{{\bf{x}}^L}} \right.} \right)} \right)}
\end{align}
where ${{\bf{p}}\left( {{{\bf{s_i}}}\left| {{{\bf{x}}^L}} \right.} \right)}$ is the probability vector f segmentation result of all pixel on the $i$th SAR chip, $n$ is the number of pixels in a SAR chip, ${s_i}$ is the segmentation labels in the form of one hot and $V$ is the number of the segmentation types. Therefore, the joint loss can  be  presented  as
\begin{align}
\label{rangereso1}
L\left( {{\bf{w,b}}} \right) = {L_r}\left( {{\bf{w,b}}} \right) + {L_s}\left( {{\bf{w,b}}} \right)
\end{align}

After the joint loss  is obtained, the optimal performance of the whole architecture could be obtained through minimizing the joint loss using backpropagation \cite{rezende2014stochastic}.

First, the total error is computed by comparing the output of the architecture with the ground truth.
\begin{align}
\label{rangereso1}
{\delta _{total}} = \sum\limits_{i = 1} {\left( {p\left( {{y_i}\left| {x^l} \right.} \right) - {y_i}} \right)}
\end{align}

Then, the error is spread from the high layer to the low layer in the architecture by computing the intermediate error of each layer. When the $l{\mathop{\rm th}\nolimits}$ layer is one convolutional layer, the intermediate error can  be calculated by
\begin{align}
\label{rangereso1}
{\delta ^l} = \left( {{{\left( {{{\bf{w}}^{l + 1}}} \right)}^T}{\delta ^{l + 1}}} \right) \odot f'\left( {{{\bf{x}}^l}} \right)
\end{align}
where $f'$ denote the 1st derivative of the ReLU, ${\delta ^l}$ denotes the intermediate error of the $l{\mathop{\rm th}\nolimits}$ convolutional layer and $ \odot $ denotes Hadamard multiplication. As for the transposed convolutional layers, the formula   is
\begin{align}
\label{rangereso1}
{\delta ^l} = \left( {{{\bf{w}}^{l + 1}}{\delta ^{l + 1}}} \right) \odot f'\left( {{{\bf{x}}^l}} \right)
\end{align}

The derivatives for updating ${{\bf{w}}^l}$ and ${b^l}$ of the $l{\mathop{\rm th}\nolimits}$ layer can  be  presented  as
\begin{align}
\label{rangereso1}
\frac{{\partial L_{\bf{w}}^l}}{{\partial {\bf{w}}_{ij}^l}} = {\bf{x}}_j^{l - 1}\delta _i^l
\end{align}
\begin{align}
\label{rangereso1}
\frac{{\partial L_b^l}}{{\partial b_i^l}} = \delta _i^l
\end{align}

This step is the same for the convolutional and transposed convolutional layers. When the backpropagation comes across the max pooling layers, only the unit with the max value in every pooling field receives the error term and the intermediate error on other units    is set as zero.

Finally, Backpropagation   updates the trainable parameters of the architecture by
\begin{align}
\label{rangereso1}
{{\bf{w}}^l} \to {{\bf{w}}^l} - lr \times \frac{{\partial L_{\bf{w}}^l}}{{\partial {{\bf{w}}^l}}}
\end{align}
\begin{align}
\label{rangereso1}
{{\bf{b}}^l} \to {{\bf{b}}^l} - lr \times \frac{{\partial L_b^l}}{{\partial {{\bf{b}}^l}}}
\end{align}
where ${{\bf{w}}^l}$ denotes the convolutional kernels of the $l{\mathop{\rm th}\nolimits}$ layer, ${b^l}$ denotes the bias of the $l{\mathop{\rm th}\nolimits}$ layer and $lr$ denotes the learning rate.

Through the process of the backpropagation, the network   gradually achieves the optimal performance, which could achieve accuracy and effective target recognition and segmentation simultaneously. Its performance   is presented  and compared in the next section.

\section{Experiments and Results}

In this section, the performance of the multi-task  deep learning framework   is evaluated. First, the information of the used dataset   is introduced in detail. Then, the steps of the data preprocessing   are described and the hyperparameter and set-up of the specific implementation of the multi-task  deep learning framework   are described. Finally, the results and comparisons of the target recognition and segmentation   are presented.

\subsection{Dataset}
\begin{figure}[H]
\centering
\includegraphics[width=5in]{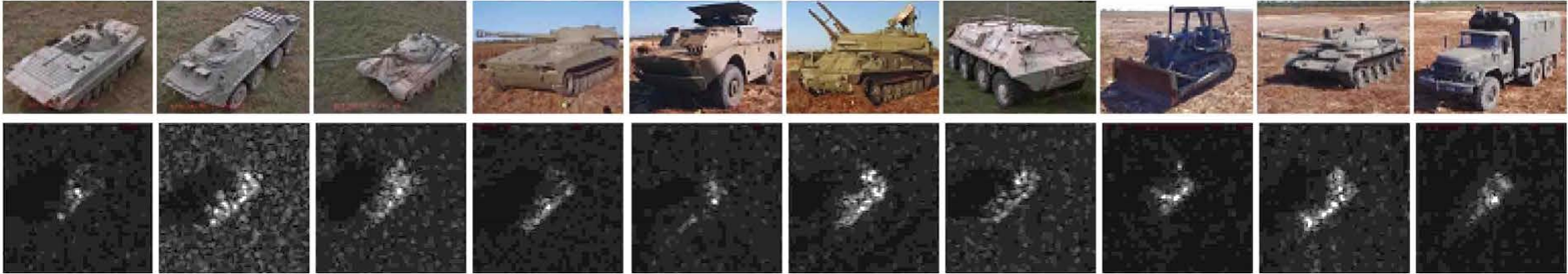}
\caption{SAR images and corresponding optical images of targets at similar aspect angels. From left to right: BMP2, BTR70, T72, 2S1, BRDM2, ZSU234, BTR60, D7, T62 and  ZIL131.}\label{Dataset}
\end{figure}

The experiment dataset used to evaluate our proposed multi-task  deep learning framework is collected from the Moving and Stationary Target Acquisition and Recognition (MSTAR) program. This dataset is released by the Defense Advanced Research Projects Agency and the Air Force Research Laboratory. The dataset is as part of the MSTAR program and collected using the Sandia National Laboratory STARLOS sensor platform \cite{ross1998standard}. As a benchmark dataset for SAR ATR performance assessment, this dataset has a significant quantity of SAR images containing different types of military vehicles and clutter images. Ten different classes of ground targets (tank, T62 and T72; rocket launcher, 2S1; truck, ZIL131; armored personnel carrier, BTR70, BTR60, BRDM2 and BMP2; air defense unit, ZSU23/4; and bulldozer,  D7) were captured as 1-ft resolution X-band SAR images with full aspect coverage (in the range of 0\degree--360\degree). They  were collected under varying operating conditions, such as different aspect angles, depression angles   and   serial numbers. As for the segmentation labels, the segmented binary labels are a precise manual marking by the tool called OpenLabeling. The SAR images and corresponding optical images of the target at similar aspect angles are depicted in Figure \ref{Dataset}.

To comprehensively assess the performance of recognition, the proposed multi-task  deep learning framework was evaluated under the standard operating condition (SOC) and extended operating condition (EOC) \cite{ross1998standard}. SOC refers to that the serial numbers and target configurations of the train and test set are the same, but with different aspects and depression angles. EOC includes three extended operating conditions: depression variant, configuration variant and version variant. As for the performance of segmentation, the proposed multi-task  deep learning framework  was  assessed with the merit of the visual and objective aspect at the same time as the assessment of the recognition performance.

\subsection{Data Preprocessing }
Before assessing the performance of the proposed multi-task  deep learning framework, data preprocessing  was employed to augment the training images and manually annotate the segmentation of the training and testing images. The specific processes are described as follows.
At first, we employed data augmentation to generate more training images \cite{perez2017effectiveness}. The numbers of the training and testing images before the data augmentation are listed in Table \ref{table1}. The training images were augmented 10 times by randomly sampling ten $88\times88$ SAR image chips from one original $128\times128$ SAR image, which ensures the central target was complete \cite{chen2016target}.
Then, the training and testing datasets of the segmentation were acquired by manual annotation using the tool named OpenLabeling. The manual annotation was based on the intensity and the contour of the target and shadow. The number of the segmentation labels was the same as the one of the original images, and, when the original images encountered the data augmentation, the segmentation labels also  went through the data augmentation in the same way. Therefore, the segmentationwas synchronous with the recognition above when the proposed network architecture was being trained or tested.
After the data preprocessing, the proposed multi-task deep learning framework could be regarded as a whole network to be trained and evaluated.

\begin{table}[H]
\renewcommand{\arraystretch}{1.3}
\caption{Number of training and testing images for SOC before the data augmentation.}
\centering
\begin{tabular}{ccccc}
\toprule
 & \multicolumn{2}{c}{\textbf{Training}} & \multicolumn{2}{c}{\textbf{Testing}} \\
\midrule
\textbf{Class} & \textbf{Depression} & \textbf{Number} & \textbf{Depression} & \textbf{Number} \\
\midrule
BMP2-9563 & 17\degree & 233 & 15\degree & 196 \\
BTR70-c71 & 17\degree & 233 & 15\degree & 196 \\
T72-132 & 17\degree & 232 & 15\degree & 196 \\
BTR60-7532 & 17\degree & 256 & 15\degree & 195 \\
2S1-b01 & 17\degree & 299 & 15\degree & 274 \\
BRDM2-E71 & 17\degree & 298 & 15\degree & 274 \\
D7-92 & 17\degree & 299 & 15\degree & 274 \\
T62-A51 & 17\degree & 299 & 15\degree & 273 \\
ZIL131-E12& 17\degree & 299 & 15\degree & 274 \\
ZSU234-d08 & 17\degree & 299 & 15\degree & 274 \\
\bottomrule
\end{tabular}
\label{table1}
\end{table}

\subsection{Network Setup}
On the basis of the proposed multi-task  deep learning framework, a specific implementation  was employed to evaluate the proposed framework for SAR ATR. The specific implement is presented in Figure \ref{network}. There are three convolutional and three max pooling layers forming the feature extractor. Two convolutional layers, one max pooling layers and one SoftMax layer are composed to accomplish the recognition task. Meanwhile, three de-convolutional layers and three convolutional layers are organized to segment the SAR images. The size of the input SAR images is $88\times88$, the stride size of every convolutional layer is $1\times1$ and the stride size for each max pooling layer is $2\times2$. Other hyper parameters in our network instances are shown in Figure \ref{network}. The weights of convolutional layers are initialized from Gaussian distributions with zero mean and a standard deviation of 0.01,   and   biases are initialized with a small constant value of 0.1. The initial learning rate is set as 0.001 and is reduced by a factor of 0.1 after 5 epochs.

\subsection{Recognition Results under SOC}
In this SOC experimental setup, the performance of the proposed architecture  was assessed on the classification of ten targets in the MSTAR dataset. The training and testing images have the same serial number, but are different in the depression angle. As listed in Table \ref{table1}, the training images  were  captured at 17\degree\ depression angle, while the testing images were captured at 15\degree\ depression angle. A summary of this experimental setup for training and testing datasets is listed in Table \ref{table1}. In Table \ref{table1}, the number of each target serial is the number of the original SAR images in MSTAR dataset before the data augmentation. The number of each class of the target after the data augmentation is 2700.

The recognition result of the proposed multi-task  deep learning is presented in Table \ref{table2}. Table \ref{table2} is a confusion matrix of ten targets, which is widely used to present the classification performance in SAR ATR \cite{mossing1998evaluation}. The numbers at the diagonal of the confusion matrix are the numbers of correct recognitions  for each target.

\begin{table}[H]
\small
\centering
\begin{spacing}{1.4}
\caption{Recognition result of the proposed MTL deep learning framework under SOC (recognition ratio 99.13\%).}
\label{table2}
\begin{tabular}{ccccccccccc}%{c|c|c|c|c|c|c|c|c|c|c|c}
\toprule
 & BMP2 & BTR70 & T72 & BTR60 & 2S1 & BRDM2 & D7 & T62 & ZIL131 & ZSU234 \\
\midrule
BMP2 & 100 & 0 & 0 & 0 & 0 & 0 & 0 & 0 & 0 & 0 \\
BTR70 & 0 & 100 & 0 & 0 & 0 & 0 & 0 & 0 & 0 & 0 \\
T72 & 0 & 0 & 100 & 0 & 0 & 0 & 0 & 0 & 0 & 0 \\
BTR60 & 0 & 0.51 & 0 & 97.95 & 0 & 1.54 & 0 & 0 & 0 & 0 \\
2S1 & 0 & 0.38 & 0.38 & 0.38 & 96.71 & 0 & 0 & 0 & 1.77 & 0.38 \\
BRDM2 & 0 & 0 & 0 & 0 & 0 & 99.64 & 0 & 0 & 0.36 & 0 \\
D72 & 0.36 & 0 & 0 & 0 & 0 & 0 & 97.81 & 0 & 0 & 1.83 \\
T62& 0 & 0 & 0.37 & 0 & 0 & 0 & 0 & 99.63 & 0 & 0 \\
ZIL131& 0 & 0 & 0 & 0 & 0 & 0 & 0 & 0 & 100 & 0 \\
ZSU234 & 0 & 0 & 0 & 0 & 0 & 0 & 0 & 0 & 0 &100 \\
\bottomrule
\end{tabular}
\end{spacing}
\end{table}

In Table \ref{table2}, the recognition ratios of BTR60, I2S1 and D7 are above 96.5\%, the recognition ratios of BRDM\_2 and T62 are above 99.5\%, and the others have achieved 100\% recognition ratio. The overall recognition ratio is 99.13\%, which is obviously satisfactory. From the recognition result, it is clear that, through the deep convolutional structure, there are some stable features extracted for the  recognition of the ten targets  among the different targets. Therefore, the proposed network architecture can achieve a satisfactory performance for the ten-target  recognition, and these results can also verify the superiority of the proposed architecture in the SOC experiment.

\subsection{Recognition Results under EOC}
In realistic battlefield situations, there is more complex target recognition in varied operation conditions, such as the variances of the depression angle and target type. Therefore, it is   necessary to assess the performance of the SAR ATR algorithm in the EOC. In this section, the stability and effectiveness of the proposed network architecture are evaluated in the variances of the depression angle, target configuration and version, which  are  denoted as EOC-D, EOC-C and EOC-V, respectively.

The SAR images are extremely sensitive to the variance of the depression angle, so it is important to evaluate the performance of the proposed network architecture at the variance of depression angle, EOC-D. However, the limitation that the MSTAR dataset only contains four targets (2S1, BRDM\_2, T-72    and   ZSU-234) which have a larger enough variance of depression angle to evaluate EOC-D. The SAR images at 17\degree\ depression angle are set as the training dataset and the corresponding SAR images at 30\degree\ depression angle are set as the testing dataset. The training dataset is generated by the same data augmentation as the SOC experiment. A summary of the training and testing dataset is listed in Table \ref{table3}. The number of each class of the training dataset   was augmented to 2700, while the number  of the training dataset   was 10,800.
\begin{table}[H]
\small
\centering
\begin{spacing}{1.4}
\caption{Number of training and testing images for EOC-D before the data augmentation.}
\label{table3}
\begin{tabular}{ccccc}
\toprule
 & \multicolumn{2}{c}{\textbf{Training}} & \multicolumn{2}{c}{\textbf{Testing}} \\
\midrule
\textbf{Class} & \textbf{Depression} & \textbf{Number} & \textbf{Depression} & \textbf{Number} \\
\midrule
2S1 & 17\degree & 299 & 30\degree & 288 \\
BRDM2 & 17\degree & 298 & 30\degree & 287 \\
T72 & 17\degree & 232 & 30\degree & 288 \\
ZSU234 & 17\degree & 299 & 30\degree & 288 \\
\bottomrule
\end{tabular}
\end{spacing}
\end{table}

The recognition performance of the proposed network architecture in the variance of depression angle is presented  in Table \ref{table5}. It can be seen that the recognition performance of the proposed multi-tasks is superior. The total recognition ratio is above 94.00\% and the recognition ratios of 2S1, BRDM-2 and ZSU-234 at 30\degree\ depression angle are higher than 93.00\%. The relatively low recognition ratio for T-72 is caused by the difference between the training and testing dataset at the depression angle and the serial number.  From the recognition performance in Table \ref{table5}, the proposed network architecture is still stable and effective when the depression angle varies greatly.

\begin{table}[H]
\small
\centering
\begin{spacing}{1.4}
\caption{Recognition result of the proposed MTL deep learning framework under EOC-D (recognition ratio 94.01\%).}
\label{table5}
\begin{tabular}{ccccc}
\toprule
 & \textbf{2S1} & \textbf{BRDM2} & \textbf{T72} & \textbf{ZSU234} \\
\midrule
2S1 & 99.31 & 0.345 & 0.345 & 0 \\
BRDM2 & 4.88 & 95.12 & 0 & 0 \\
T72 & 11.81 & 0 & 88.19 & 0 \\
ZSU234 & 1.45 & 0 & 5.45 & 93.10 \\
\bottomrule
\end{tabular}
\end{spacing}
\end{table}

The performance  of the proposed network architecture  with  the variance of target configuration and version (EOC-C and EOC-V)  was also evaluated. Limited by the difficulty of acquiring the SAR images of different configurations and versions of targets, the training datasets for EOC-C and EOC-V could only be set as four targets (BMP-2, BRDM\_2, BTR-70    and   T-72) at 17\degree\ depression angle and the testing datasets are set as the corresponding SAR images of the targets with different configurations and versions. The numbers of the training data of the four targets before the data augmentation are listed in Table \ref{paraa}, and the testing datasets are listed in Tables \ref{table6} and \ref{table7}. The number  of each class of the four targets in the training dataset    was augmented to 2700.  In  Tables \ref{paraa} and \ref{table6}, there are two different configurations of BMP2 and five different configurations of T72 captured at 17\degree\ and 15\degree\ depression angles to evaluate the recognition performance under the EOC of the target configuration varieties.  In Tables \ref{paraa} and \ref{table7}, it can be seen that the testing dataset for EOC-V has four different serial types of T72 from the training dataset, which are captured at 17\degree\ and 15\degree\ depression angles and utilized to evaluate the recognition performance of the proposed multi-task deep learning framework under the EOC of the target version varieties.

\begin{table}[H]
\small
\centering
\begin{spacing}{1.4}
\caption{Number of training images for EOC-C and EOC-V.}
\label{paraa}
\begin{tabular}{ccc}
\toprule
 & \multicolumn{2}{c}{\textbf{Training}} \\
\midrule
\textbf{Class} & \textbf{Depression} & \textbf{Number} \\\midrule
BMP2-9563 & 17\degree & 233 \\
BTR70-c71& 17\degree & 233 \\
T72-132 & 17\degree & 232 \\
BRDM2-E71 & 17\degree & 256 \\
\bottomrule
\end{tabular}
\end{spacing}
\end{table}

\begin{table}[H]
\small
\centering
\begin{spacing}{1.4}
\caption{Number of testing images for EOC-C.}
\label{table6}
\begin{tabular}{cccc}
\toprule
\textbf{Class} & \textbf{Serial No.} & \textbf{Depression} & \textbf{Number} \\
\midrule
\multirow{2}{*}{BMP2}
 &9566 & 15\degree, 17\degree & 428 \\
\cline{2-4}
 &C21 & 15\degree, 17\degree & 429 \\
\multirow{5}{*}{T72}
 &812& 15\degree, 17\degree & 426 \\
\cline{2-4}
 &A04 & 15\degree, 17\degree & 573 \\
\cline{2-4}
 &A05 & 15\degree, 17\degree & 573 \\
\cline{2-4}
 &A07 & 15\degree, 17\degree & 573 \\
\cline{2-4}
 &A10 & 15\degree, 17\degree & 567 \\
\bottomrule
\end{tabular}
\end{spacing}
\end{table}
\unskip
\begin{table}[H]
\small
\centering
\begin{spacing}{1.4}
\caption{Number of testing images for EOC-V.}
\label{table7}
\begin{tabular}{cccc}
\toprule
\textbf{Class} & \textbf{Serial No.} & \textbf{Depression} & \textbf{Number} \\
\midrule
\multirow{5}{*}{T72}
 &S7 & 15\degree, 17\degree & 419 \\
\cline{2-4}
 &A32 & 15\degree, 17\degree & 572 \\
\cline{2-4}
 &A62 & 15\degree, 17\degree & 573 \\
\cline{2-4}
 &A63 & 15\degree, 17\degree & 573 \\
\cline{2-4}
 &A64 & 15\degree, 17\degree & 573 \\
\bottomrule
\end{tabular}
\end{spacing}
\end{table}

The recognition performance of the proposed network architecture in EOC-C is presented  in Table \ref{table8}. The recognition performance of the proposed network architecture is 98.36\% for the variance of target configuration. It can be proved that the proposed network architecture has the  ability to recognize the targets with different configurations. As for the recognition performance in EOC-V, which is presented  in Table \ref{table9}, the recognition ratio has reached 99.21\% for the  five versions of T72. The proposed network architecture is       resilient to the variance of the target version.

From the four experiment results of SOC, EOC-D, EOC-C and EOC-V, the proposed network architecture has    obtained  superior recognition performance. It demonstrates that the proposed multi-task  deep learning framework has the  ability  to extract  optimal and effective target features from SAR images, which are also resilient to the variances of the depression angle, target configuration and version.

\begin{table}[H]
\small
\centering
\begin{spacing}{1.4}
\caption{Recognition result of the proposed MTL deep learning framework under EOC-C (recognition ratio 98.36\%).}
\label{table8}
\begin{tabular}{ccccc}
\toprule
 \textbf{Class} & \textbf{BMP2} & \textbf{BRDM2} & \textbf{BTR70} & \textbf{T72} \\
\midrule
BMP2sn-9566 & 96.93 &0.23 & 1.64 & 4.21 \\
BMP2sn-c21 & 96.04 & 0.47 & 0.47 & 3.03 \\
T72sn-812 & 0.00 & 0.47 & 0.47 & 99.06 \\
T72-A04 & 0.17 & 0.17 & 0.00 & 99.65 \\
T72-A05 & 0.00 &0.00 & 0.00 & 100.00 \\
T72-A07 & 0.17 &0.00& 0.00 &99.83 \\
T72-A10 & 0.00 &0.00 & 0.00 &100.00 \\
\bottomrule
\end{tabular}
\end{spacing}
\end{table}
\unskip
\begin{table}[H]
\small
\centering
\begin{spacing}{1.4}
\caption{Recognition result of the proposed MTL deep learning framework under EOC-V (recognition ratio 99.21\%).}
\label{table9}
\begin{tabular}{ccccc}
\toprule
 \textbf{Class} & \textbf{BMP2} & \textbf{BRDM2} & \textbf{BTR70} & \textbf{T72} \\
\midrule
T72sn-s7 & 1.19 & 0.23 & 0.23 & 98.33 \\
T72-A32 & 0.00 & 0.00 & 0.00 & 100.00 \\
T72-A62 & 1.57 &0.17 & 0.00 & 98.25 \\
T72-A63 & 0.17 &0.00& 0.35 &99.48 \\
T72-A64 & 0.00 &0.00 & 0.00 &100.00 \\
\bottomrule
\end{tabular}
\end{spacing}
\end{table}

\subsection{Results of SAR Target Segmentation}

\begin{figure}[H]
\centering
%\begin{center}
\subfigure[]{\label{1.1}\includegraphics[scale=0.25]{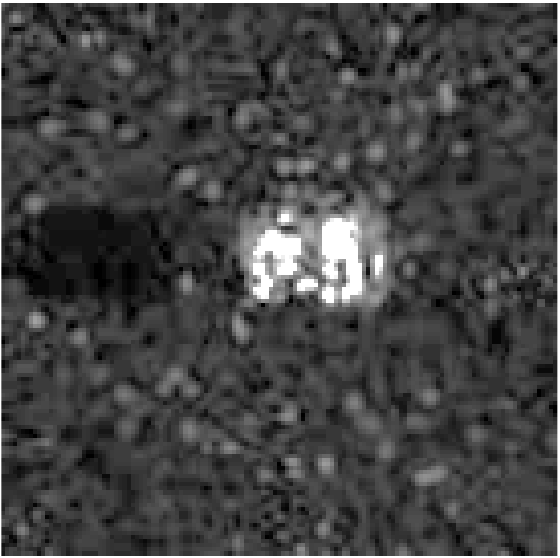}} \subfigure[]{\label{1.2}\includegraphics[scale=0.2475]{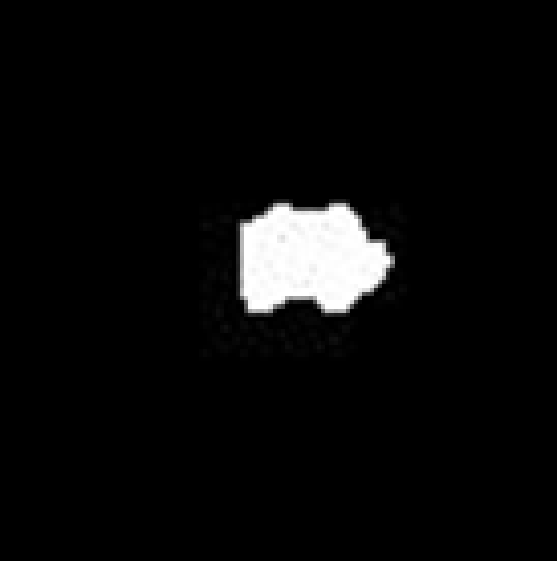}} \subfigure[]{\label{1.3}\includegraphics[draft=false,scale=0.25]{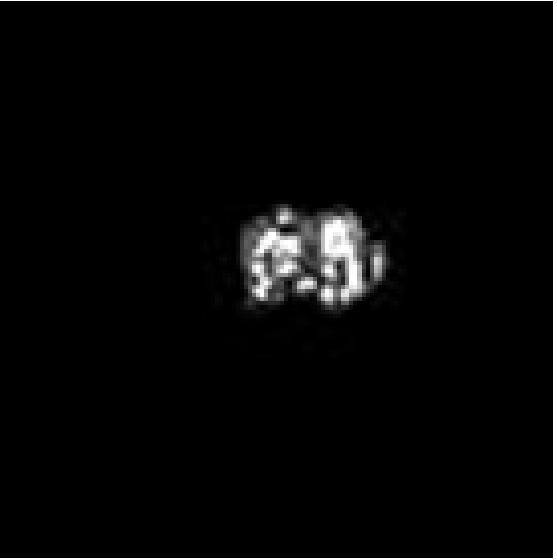}}\\
%\end{center}
\caption{Some examples of the segmentation labels for ten targets: (\textbf{a}) SAR image; (\textbf{b}) segmentation ground truth; and  (\textbf{c}) masked original image}
\label{figure4}
\end{figure}

As mentioned above, the segmentation of the targets in SAR images not only is able to obtain   more refined structural features in morphology, but also could obtain the semantic information in the pixel level. Some examples of the segmentation labels for targets are presented  in Figure \ref{figure4}. In Figure \ref{figure4}, the left image is the original image in the MSTAR dataset and the middle one is the segmentation ground truth. The right image is the original image masked by the ground truth, which    is denoted as the masked original image.

\begin{figure}[H]
\centering
%\begin{center}
\subfigure[]{\label{1.1}\includegraphics[scale=0.75]{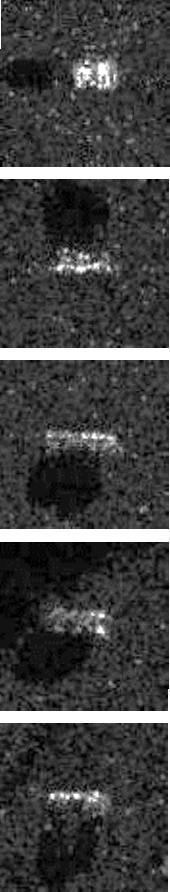}} \subfigure[]{\label{1.2}\includegraphics[scale=0.75]{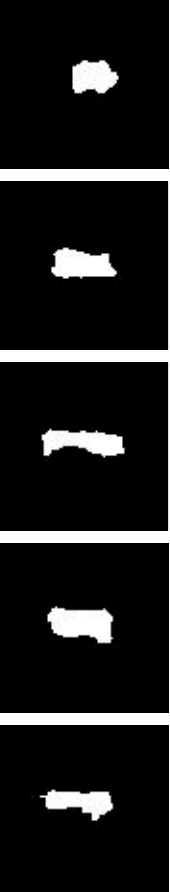}} \subfigure[]{\label{1.3}\includegraphics[scale=0.75]{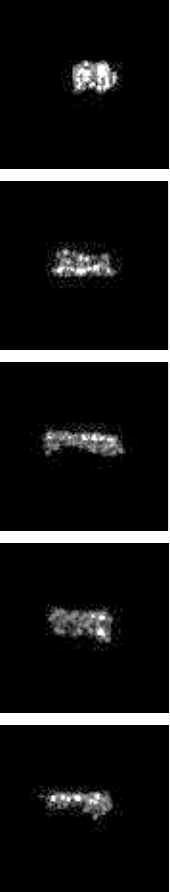}} \subfigure[]{\label{1.3}\includegraphics[scale=0.75]{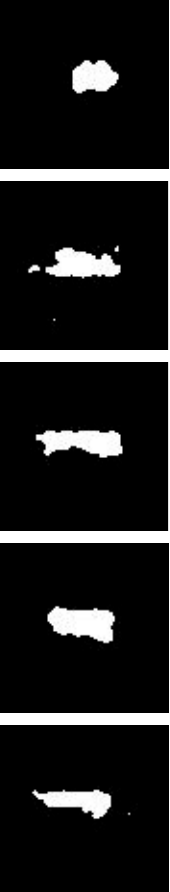}} \subfigure[]{\label{1.3}\includegraphics[scale=0.75]{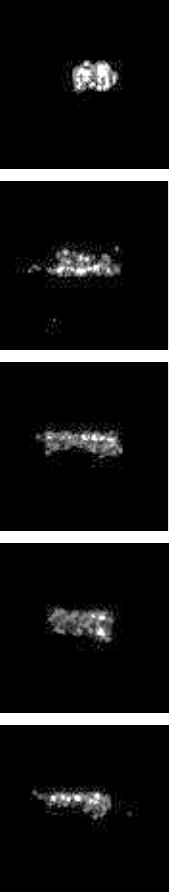}}\\
%\end{center}
\caption{Some segmentation results of the proposed network architecture for different targets: (\textbf{a}) original SAR image; (\textbf{b}) segmentation ground truth; (\textbf{c}) masked original image; (\textbf{d}) segmentation results; and  (\textbf{e}) masked segmentation results.}
\label{figure5}
\end{figure}

To present the segmentation results visually,     some segmentation results of the proposed network architecture for different targets are shown  in Figure \ref{figure5}. The first three columns are the original SAR images from the MSTAR dataset, the segmentation ground truth and their corresponding masked original SAR images, respectively. The fourth column is the segmentation results of the proposed multi-task  deep learning framework.  The last column is the original SAR images masked by the segmentation results. It can be seen that the segmentation results of the proposed multi-task  deep learning framework are quite close to the segmentation ground truth in the morphological contour. It can be concluded  that the proposed network architecture can segment precisely when the contour and intensity of the targets are varying.

To evaluate the segmentation results more objectively, the pixel accuracy of the segmentation results is employed, which   evaluates the accuracy of segmenting the targets from the background. The pixel accuracy is calculated as follows.
\begin{align}
\label{rangereso1}
{P_{pa}} = sum\left( {{P_p}} \right)/\left( {{P_a}} \right)
\end{align}
where ${P_{pa}}$ is the pixel accuracy, ${P_p}$ is the correct predicted pixel and ${P_a}$ is the total pixels in one SAR image.  It means that the higher the pixel accuracy is, the better the performance is.

The pixel accuracy of the proposed multi-task  deep learning framework is presented in the form of a confusion matrix in Table \ref{table10}.  In Table \ref{table10}, the accuracy for the target or background is above 98.00\% and the overall accuracy of the segmentation is higher than 99.00\%. From the quantitative analyses, it is quite clear that the proposed network architecture has the  ability to segment  the targets from the backgrounds precisely and effectively.

From the evaluations of the performance of the target recognition and segmentation, it can be proved that, through the deep learning structure of multiple convolutional layers and the multi-task  framework design of the encoder and two sub-decoders, the proposed multi-task  deep learning framework can achieve the target recognition and segmentation accurately and effectively and finish those two tasks simultaneously with only one system.

\begin{table}[H]
\small
\centering
\begin{spacing}{1.4}
\caption{Pixel accuracies for the targets and the backgrounds (pixel accuracy 99.03\%).}
\label{table10}
\begin{tabular}{ccc}
\toprule
 \textbf{Pixel Accuracy} & \textbf{Target} & \textbf{Background} \\
\midrule
Target & 98.92 & 1.08 \\
Background & 0.87 & 99.13 \\
\bottomrule
\end{tabular}
\end{spacing}
\end{table}

\subsection{Comparison of Performance of Segmentation and Recognition }

In this section, we compare our proposed algorithm with other algorithms in recognition and segmentation. For recognition,   seven SAR ATR algorithms are considered:  support vector machine (SVM) \cite{srinivas2014sar}, adaptive boosting (AdaBoost) \cite{srinivas2014sar} IGT \cite{srinivas2014sar}, CGM \cite{o2001sar}, two DCNNs and gcForest \cite{zhang2020sar}. SVM and AdaBoost, both traditional algorithms, IGT, based on the probabilistic graphical model, and the  two DCNNs \cite{ding2016convolutional,morgan2015deep} are     state of the art  in SAR ATR, while gcForest is recently published. For segmentation,   two other algorithms are considered, namely  Maximum Between-Class Variance (Otsu Method) \cite{kaur2012comparative} and Canny edge detector (Canny) \cite{al2012evaluation}, which are traditional algorithms for segmentation in SAR images.

For recognition performance comparison, we compare those algorithms with our proposed algorithm in terms of the recognition performance. The recognition performances are listed in Table \ref{table11} under SOC and EOC.  In Table \ref{table11}, the performance of our proposed algorithm is better than other algorithms under SOC and has significant improvement under EOC. Therefore,  can be  concluded  that our proposed algorithm is superior to other algorithms in recognition performance.
\begin{table}[H]
\small
\centering
\begin{spacing}{1.4}
\caption{Recognition performance for various methods.}
\label{table11}
\begin{tabular}{cccc}
\toprule
 \textbf{Methods} & \textbf{SOC} & \textbf{EOC-D} & \textbf{EOC-V} \\
\midrule
SVM \cite{srinivas2014sar} & 90.00\% & 75.00\% & 81.00\% \\
AdaBoost \cite{srinivas2014sar} & 92.00\% & 78.00\% & 82.00\% \\
IGT \cite{srinivas2014sar} & 95.00\% &80.00\% & 85.00\% \\
CGM \cite{o2001sar} & 97.00\% &79.00\%&80.00\% \\
DCNN \cite{morgan2015deep} & 92.30\% & $-$ & $-$ \\
DCNN \cite{ding2016convolutional} & 94.56\% &$-$ & $-$ \\
gcForest \cite{zhang2020sar} & 96.70\% & $-$ & $-$ \\
Proposed method & 99.13\% &94.01\% &99.21\% \\
\bottomrule
\end{tabular}
\end{spacing}
\end{table}

\begin{figure}[H]
\centering
%\begin{center}
\subfigure[]{\label{1.1}\includegraphics[scale=0.75]{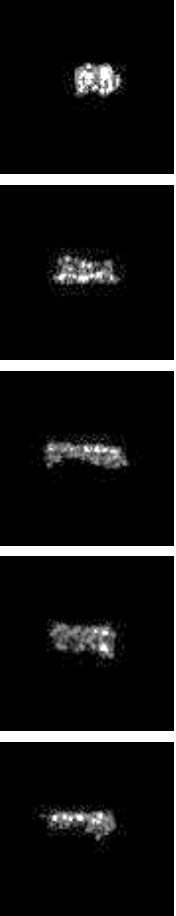}} \subfigure[]{\label{1.2}\includegraphics[scale=0.75]{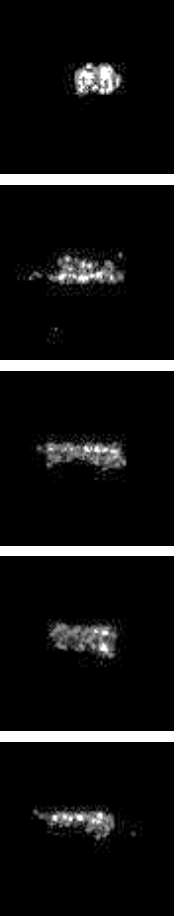}} \subfigure[]{\label{1.3}\includegraphics[scale=0.75]{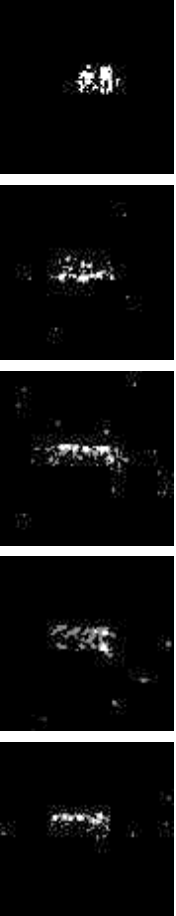}} \subfigure[]{\label{1.3}\includegraphics[scale=0.75]{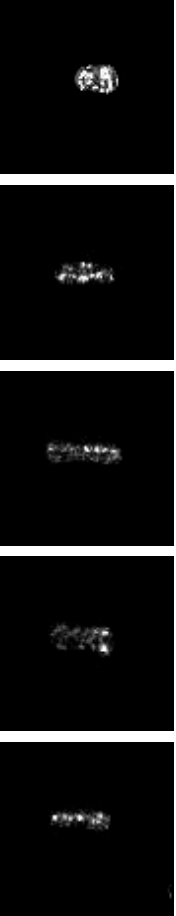}}\\
%\end{center}
\caption{Some segmentation results of the proposed network architecture for different targets present: (\textbf{a}) masked original SAR images; (\textbf{b}) masked segmentation results of the proposed method; (\textbf{c}) masked segmentation results of Otsu; and   (\textbf{d}) masked segmentation results of Canny.}
\label{figure6}
\end{figure}

For segmentation performance comparison,     some segment images of different SAR images using Otsu, Canny    and   our proposed algorithm are shown in Figure \ref{figure6}.  In Figure \ref{figure6}, it is obvious that our proposed algorithm has better performance than other algorithms when the image intensity varies and the contour of images is complicated. At the same time, the pixel accuracies of Otsu, Canny and our proposed algorithm are listed in Table \ref{table12}.  In  Table \ref{table12}, it is clear that the proposed multi-task  deep learning framework has higher pixel accuracy than the other algorithms. From the comparisons of the segmentation above,  it can be  concluded  that the proposed multi-task  deep learning framework could obtain more accurate segmentation at both the overall contour and local details of the targets.

\begin{table}[H]
\small
\centering
\begin{spacing}{1.4}
\caption{Pixel accuracies of Otsu, Canny and our proposed algorithm.}
\label{table12}
\begin{tabular}{cccc}
\toprule
\textbf{Methods} & \textbf{Pixel Accuracy of Target} & \textbf{Pixel Accuracy of Background} & \textbf{Pixel Accuracy} \\
\midrule
Otsu & 58.17\% & 88.35\% & 73.26\% \\
Canny & 79.12\% & 90.13\% & 85.12\% \\
Proposed & 98.92\% & 99.13\% & 99.03\% \\
\bottomrule
\end{tabular}
\end{spacing}
\end{table}
\unskip

From the above all the contrast experiments, it is clear that, through the deep learning structure and the   multi-task capability, the proposed multi-task deep learning framework not only could extract the optimal effective target feature to achieve the accurate robust recognition, but also could obtain the overall contour and local details of the targets to achieve elaborate segmentation at the same time as the recognition.
All the evaluations and the contrast experiments   verify that our proposed algorithm has the superiority in both recognition and segmentation with the capability of   simultaneous target recognition and segmentation.

\section{Conclusions}
When deep learning meets multi-task learning, multi-task learning will acquire the capability of adaptive feature learning and powerful feature representation to promote the performances of multiple tasks simultaneously in SAR ATR. Hence, we propose  a novel multi-task deep learning framework to obtain accurate category and precise shape of the targets simultaneously. With an elaborately designed encoder, the optimal image features are extracted from different scales to represent the overall contour and local details of the target. With employing these extracted features adaptively and optimally to meet the different feature demands of the recognition and segmentation, the task-specific decoder achieves superior performance in   terms of recognition and segmentation simultaneously. Extensive experiments  were  carried out on the MSTAR dataset,   and   the results show clearly that the proposed framework not only achieves higher recognition performance than existing SAR ATR methods in SOC and EOCs, but also obtains more precise and stable segmentation performance than other segment methods. With the sufficient target attributes extracted by the proposed multi-task framework, it could make some contributions to the practical application of SAR ATR systems.

\vspace{6pt}
\authorcontributions{Conceptualization, C.W. and J.P.; methodology, C.W.; software, J.P.; validation, C.W., Z.W., J.P. and J.W.; formal analysis, C.W. and J.W.; investigation, H.Y. and Y.Huang; resources, C.W. and Y.Huang; data curation, C.W. and Y.Huang; writing---original draft preparation, C.W.; writing---review and editing, C.W. and Y.G.; visualization, C.W., Z.W. and Y.Huang; supervision, H.Y., Y.G. and J.Y.; project administration, C.W. and H.Y.; and funding acquisition, J.P. and H.Y. All authors have read and agreed to the published version of the manuscript.}

%%%%%%%%%%%%%%%%%%%%%%%%%%%%%%%%%%%%%%%%%%
\funding{This research was funded by the National Natural Science Foundation of China under grant numbers 61901091 and 61671117.}

%%%%%%%%%%%%%%%%%%%%%%%%%%%%%%%%%%%%%%%%%%
\conflictsofinterest{The authors declare no conflict of interest.}

%%%%%%%%%%%%%%%%%%%%%%%%%%%%%%%%%%%%%%%%%%
\reftitle{References}
\externalbibliography{yes}

\bibliography{refe,ref_self}

% Please provide either the correct journal abbreviation (e.g. according to the ��List of Title Word Abbreviations�� http://www.issn.org/services/online-services/access-to-the-ltwa/) or the full name of the journal.


%=====================================
% References, variant B: internal bibliography
%=====================================

% The following MDPI journals use author-date citation: Arts, Econometrics, Economies, Genealogy, Humanities, IJFS, JRFM, Laws, Religions, Risks, Social Sciences. For those journals, please follow the formatting guidelines on http://www.mdpi.com/authors/references
% To cite two works by the same author: \citeauthor{ref-journal-1a} (\citeyear{ref-journal-1a}, \citeyear{ref-journal-1b}). This produces: Whittaker (1967, 1975)
% To cite two works by the same author with specific pages: \citeauthor{ref-journal-3a} (\citeyear{ref-journal-3a}, p. 328; \citeyear{ref-journal-3b}, p.475). This produces: Wong (1999, p. 328; 2000, p. 475)

%%%%%%%%%%%%%%%%%%%%%%%%%%%%%%%%%%%%%%%%%%
%% optional

%% for journal Sci
%\reviewreports{\\
%Reviewer 1 comments and authors�� response\\
%Reviewer 2 comments and authors�� response\\
%Reviewer 3 comments and authors�� response
%}

%%%%%%%%%%%%%%%%%%%%%%%%%%%%%%%%%%%%%%%%%%
\end{document}